\documentclass[a4paper,11pt]{article}

\usepackage{pos}

\title{Constraints on the Population of Common Sources of Gravitational Waves and High-Energy Neutrinos with IceCube During the Third Observing Run of the LIGO and Virgo Detectors}

\ShortTitle{Constraints on GWHEN sources from O3}

\author{Do\u{g}a Veske$^{1,2*}$,
Zsuzsa M\'arka$^{2}$, 
Albert Zhang$^{3}$ on behalf of the IceCube Collaboration$^a$, the LIGO Scientific Collaboration, the Virgo Collaboration, and the KAGRA Collaboration \\{\normalsize \normalfont
{$^{1}$ \itshape Fizik B\"ol\"um\"u, Orta Do\u{g}u Teknik \"Universitesi, \c{C}ankaya, Ankara 06800, Turkey}\\
{$^{2}$ \itshape Columbia Astrophysics Laboratory, Columbia University in the City of New York, New York, NY 10027, USA}\\
{$^{3}$ \itshape Department of Physics, Columbia University in the City of New York, New York, NY 10027, USA}\\
{$^a$ IceCube Collaboration (a complete list of authors can be found at the end of the proceedings)}\\
$^*$ Presenter}\\}

\emailAdd{veske@metu.edu.tr}
\emailAdd{zsuzsa@astro.columbia.edu}
\emailAdd{acz2122@columbia.edu}

\abstract{

The discovery of joint sources of high-energy neutrinos and gravitational waves has been a primary target for the LIGO, Virgo, KAGRA, and IceCube observatories. The joint detection of high-energy  neutrinos and gravitational waves would provide insight into cosmic processes, such as progenitor dynamics and outflows. The joint detection of multiple cosmic messengers can also elevate the significance of the observation when some or all of the constituent messengers are sub-threshold, not significant enough to declare their detection individually. Leveraging data from the  LIGO, Virgo, and IceCube observatories, we conducted an archival investigation of sub-threshold multimessenger events. Complementing previous analyses, we used minimal assumptions to search for common sources of sub-threshold gravitational-wave and high-energy neutrino candidates during the third observing run (O3) of the Advanced LIGO and Advanced Virgo detectors. Our search did not identify  significant joint sources. We therefore derive constraints on the rate density of joint sources for each compact binary merger population as a function of the energy emitted in neutrinos. Only a fraction of the gravitational-wave sources emit neutrinos, if the neutrino emission has high bolometric energy ($>10^{52}$ to $10^{54}$ erg).

\vspace{4mm}
}

\ConferenceLogo{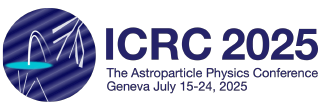}

\FullConference{39th International Cosmic Ray Conference (ICRC2025)\\
 15–24 July 2025\\
Geneva, Switzerland\\}

\begin{document}

\maketitle

\section{Introduction}
\label{sec:intro}
Traditionally astrophysical phenomena are observed via single messengers. These observations provide partial information about the sources, only about the processes that emit the observed messenger. A complete picture of many sources can only be depicted by multimessenger observations. Since the discovery of gravitational waves (GW)~\citep{Abbott_2016}, about 90 confident compact binary mergers have been observed by end of the third observing run of the LIGO~\citep{2015CQGra..32g4001L} and Virgo~\citep{2015CQGra..32b4001A} detectors (O3)~\citep{2023PhRvX..13d1039A}, including the first two runs O1 and O2. However, except for the binary neutron star (BNS) merger in 2017~\citep{2017PhRvL.119p1101A}, no other messenger has been observed from them. From the multimessenger observations of that BNS merger GW170817~\citep{2017ApJ...848L..12A}, we have learned a lot about the processes in BNS mergers. In fact we were informed in different areas as well, from nuclear physics to cosmology. Multimessenger observations of more GW events can help us answer many questions. For instance, how binary black holes (BBH) form is still a mystery. If counterparts are observed from BBH mergers, they can hint about the formation regions such as near active galactic nuclei.

Here we are searching for high-energy neutrino (HEN) emission from GW events within a time window of $\pm500$~s~\citep{2011APh....35....1B}. We use the HEN data of the IceCube Neutrino Observatory~\citep[IceCube hereafter;][]{Aartsen_2017,Aartsen_2024} and the GW data from the LIGO and Virgo detectors. Discovery of HEN emission from GW events can teach us not only about the GW events; but also about the unknown sources of the observed astrophysical diffuse HEN flux. The searches for joint GW and HEN events have only yielded coincidences that are not statistically significant~\citep{2020ApJ...898L..10A,2022icrc.confE.950V,Thwaites:20236w,2023ApJ...944...80A}. The absence of statistically significant coincidences among approximately 90 events highlights the inherent difficulty of finding true coincidences. We should use powerful search methods along with datasets as large as possible~\citep{2021ApJ...908..216V}. In this study we searched for HEN emission with the Low-Latency Algorithm for Multimessenger Astrophysics (LLAMA) pipeline~\citep{2019arXiv190105486C} from the candidate GW events from O3 that have a false alarm rate up to 2/day, of which there are 2210. The analysis of candidate GW events, most of which have high probability of being noise, was done via the construction of a Bayesian test statistic that accounts for the probability of them being noise~\citep{2019PhRvD.100h3017B}. The strength of this study compared to the previous analyses is the extended dataset. To search for a population of joint GW and HEN events in this population we employ an optimized test statistic for population level studies. 

In our search, we did not identify a significant population of jointly GW and HEN emitting events. We use their absence in our results to constrain their presence in the universe. We introduce our datasets in Sec. \ref{sec:data}. In Sec. \ref{sec:method} we describe the methods we use. We present our results and briefly discuss them in Sec. \ref{sec:res}.

\section{Data}
\label{sec:data}
Our analysis has two datasets: one for candidate GW events and one for HEN events. 
\subsection{High-energy neutrino data}
IceCube is a cubic kilometer neutrino detector located at the South Pole. It is composed of 5160 digital optical modules deployed in clear ice, looking for Cherenkov radiation from neutrino interactions. The observed events are classified into track and cascade categories. Tracks include the interactions resulting in a muon or anti-muon that traverses a straight path, whereas cascades include the showers occurred in ice. The directional reconstruction of the original neutrino or anti-neutrino is  $\lesssim1^\circ$ in the case of tracks, and $\sim10^\circ$ for cascades. For astronomical associations with specific sources the directional accuracy is important. Hence, in our search we use the Gamma-ray Follow-Up (GFU) dataset, which consists of track events~\citep{2016JInst..1111009I}. IceCube maintains nearly continuous operation with over 99\% duty cycle~\citep{Aartsen_2017,Aartsen_2024}. This dataset has $\sim6.4$~mHz event rate. The energy of the neutrinos spans from a few hundred GeV to a few PeV. The sensitivity to the Northern Hemisphere is higher thanks to the Earth blocking the high flux of atmospheric muons, where most of the low-energy ($\mathcal{O}$(1)~TeV) portion of the dataset comes from.

We use the detection time, reconstructed direction and uncertainty, and the reconstructed energy of the neutrinos as inputs in our analysis.
\subsection{Gravitational wave data}
The LIGO and Virgo GW detectors are Michelson type interferometers. Their arm lengths are 4~km and 3~km, respectively. These lengths are effectively increased by a factor of hundreds with Fabry-P\'erot cavities. There are two LIGO detectors: one at Hanford, Washignton and one at Livingston, Louisiana in the United States. Virgo is located at Cascina, Italy. O3 lasted from April 2019 to March 2020. The detectors had duty cycles of around 70-80\% during O3. LIGO Hanford had a BNS range of $\sim110$~Mpc, LIGO Livingston had $\sim135$~Mpc, and Virgo had $\sim50$~Mpc. The collected data have been analyzed with several different analyses. We use the public results of the template based search pipelines GstLAL, PyCBC-BBH, PyCBC-broad, and MBTA for compact binary coalescences (CBC) up to a false alarm rate of 2/day. This dataset has 2210 candidate GW events including 79 confident detections. 

We use the event time, reconstructed sky localization and luminosity distance, signal to noise ratio, and the estimated probability of being astrophysical ($p_{\rm astro}$) as inputs in our analysis.
\section{Method}
\label{sec:method}
Our analysis has two parts: analysis of coincidences from individual events, and population inference and constraints from the search results.
\subsection{Individual coincidences}
We employ the LLAMA pipeline in our search. The search is based on assigning frequentist significances. The test statistic is constructed from four hypotheses:
\begin{itemize}
    \item $H_{\rm s}$: signal hypothesis, in which the HEN and GW come from the same astrophysical source
    \item $H_{\rm c}^{\rm GW}$: GW coincidence hypothesis, in which the GW event is astrophysical but the neutrino is not
    \item $H_{\rm c}^{\nu}$: HEN coincidence hypothesis, in which the HEN is astrophysical but the GW is not
    \item $H_{\rm n}$: noise hypothesis, in which both the HEN and GW are not astrophysical
\end{itemize}
We combine the likelihoods for $H_{\rm c}^{\rm GW}$, $H_{\rm c}^{\nu}$, and $H_{\rm c}^{\nu}$ using their estimated occurrence rates as Bayesian priors. In the end, our test statistic for inputs $\mathbf{x}$ is
\begin{equation}
    {\rm TS}\left(\mathbf{x}\right)=\frac{P\left(H_{\rm s}\right)P\left(\mathbf{x}|H_{\rm s}\right)}{P\left(H_{\rm c}^{\rm GW}\right)P\left(\mathbf{x}|H_{\rm c}^{\rm GW}\right)+P\left(H_{\rm c}^{\nu}\right)P\left(\mathbf{x}|H_{\rm c}^{\nu}\right)+P\left(H_{\rm n}\right)P\left(\mathbf{x}|H_{\rm n}\right)}
\end{equation}
Theoretically there could be a fifth hypothesis for astrophysical HEN and GW events coming from different sources; but the rate of such a coincidence is extremely low and we neglect it. The details of the likelihoods can be found in Ref.~\citep{2019PhRvD.100h3017B}. We search for HEN emission $\pm500$~s around the mergers~\citep{2011APh....35....1B}. Considering the 6.4~mHz event rate in the GFU dataset, we expect on average 6.4 HEN candidates temporally coincident but unrelated to the GW event.

The background distribution for finding frequentist significances is obtained by creating false coincidences. This is done by scrambling the detection time of our neutrino dataset and analyzing resulting coincidences. We use a single background distribution for obtaining $p$-values.
\subsection{Population inference}
The individual event analysis might be useful when analyzing small number of events. However, when a large set of events is analyzed and the presence of a population is sought in them, it becomes a weaker approach. The large number of events should be accounted for by a trials correction. The widely used Bonferroni correction which effectively multiplies the obtained $p$-values by the number of trials. Another approach could be only considering the lowest $p$-value event. However, both of these approaches are suboptimal in this case. 

When the sought signal has a rate much lower than the unrelated events, the optimal test statistic for finding a population becomes the \emph{sum} of individual test statistics. Hence we define the population test statistic
\begin{equation}
    {\rm TS^{pop}}=\sum_i{\rm TS(\mathbf{x_i})}.
\end{equation}
To obtain the background distribution for this population test statistic, we construct datasets consisting of false coincidences. This was done by repeatedly sampling 2210 (the number of GW events in our analysis) events from our individual event background distribution. Each such realization with 2210 events becomes one dataset. Collection of such many datasets and finding ${\rm TS^{pop}}$ for each determines our population background distribution.

The constraints on the rate density of GW and HEN emitting events (GWHEN events) from ${\rm TS^{pop}}$ is obtained by the standard Neyman construction. An upper limit corresponds to a rate density that would produce a higher test statistic than the observed test statistic with a probability equal to the confidence level. This probability depends on the number of GWHEN events in the dataset. For a given rate density $\dot{n}$ and observation duration $T_{\rm obs}$, the expected number of GWHEN signal events is 
\begin{equation}
    \langle N_{\rm GWHEN}\rangle= \dot{n}\times T_{\rm obs}\int 2\pi f(D_\mathrm{L},z) p^\nu_{\rm det}(E_{\nu},D_\mathrm{L},\delta)p^{\rm GW}_{\rm det}(\mathcal{S},D_\mathrm{L})\cos\delta {\rm d}\delta {\rm d}D_\mathrm{L}.
\end{equation}
where $p^\nu_{\rm det}(E_{\nu},D_\mathrm{L},\delta)$ and $p^{\rm GW}_{\rm det}(\mathcal{S},D_\mathrm{L})$ are the detection probabilities of HENs and GWs respectively from an event with total HEN emission energy $E_{\nu}$ and the specific GW event type $\mathcal{S}$, as a function of luminosity distance $D_\mathrm{L}$ and declination $\delta$. The GW event type refers to the components of the merger such as BNS, BBH, or neutron star black hole binary (NSBH). $f(D_\mathrm{L},z)$ gives the distribution of GWHEN events as a function of distance and redshift $z$ considering the cosmological expansion as
\begin{equation}
    f(D_\mathrm{L},z)=(1+z)^{-1}\frac{{\rm d}V_{\rm c}}{{\rm d}D_\mathrm{L}}=\frac{c}{(1+z)^3H_0\sqrt{\Omega_m(1+z)^3+\Omega_\Lambda}}\frac{{\rm d}z}{{\rm d}D_\mathrm{L}}D_\mathrm{L}^2,
    \label{eq:cosmo}
\end{equation}
assuming a constant rate in comoving volume $V_{\rm c}$ where $c$ is the speed of light, $H_0=67.9$~km~s$^{-1}$~Mpc$^{-1}$ is the Hubble constant, and $\Omega_{\rm m}=0.3065$ and $\Omega_\Lambda=0.6935$ are the energy densities of matter and dark energy~\citep{2016A&A...594A..13P}.

In order to know the ${\rm TS^{pop}}$ distribution under different number of GWHEN events, which is necessary for deriving upper limits, we performed injections of GW events and HEN. The GW event injections were done and localized via BAYESTAR~\citep{2016PhRvD..93b4013S}. We injected BBH, NSBH, and BNS merger events separately. The astrophysical properties of them were chosen according to the most recent estimates after O3~\citep{PhysRevX.13.011048}. For simulating the HEN emission, we injected a single Monte Carlo generated neutrino from the injection point of the GW on the sky. The individual energies of the HEN ($\epsilon$) were chosen according to an $\epsilon^{-2}$ spectrum, which is the expected spectrum for the neutrinos from cosmic rays accelerated by Fermi acceleration~\citep{1983RPPh...46..973D,Kurahashi_2022}.

\section{Results and discussion}
\label{sec:res}
From our analysis of 2210 events individually, the most significant coincidence has a $p$-value of $3.8\times10^{-4}$ (0.84 after Bonferroni correction). It is from a candidate BNS merger event with 0.997 probability of originating from noise, detected at the GPS time 1262142545.615. The candidate GW event's mean reconstructed luminosity distance is 295 Mpc. The GFU event that produces most of the significance has a muon proxy energy of 2.4 TeV. Its position is right above the equator. Although that direction is a high-sensitivity region for IceCube, its energy is not particularly outstanding in the GFU sample. It was detected 222~s before the candidate GW event.

When we did the population analysis, we obtained a $p$-value of 0.22 for the whole dataset. Clearly there is no significant sign of a GWHEN population in the analyzed dataset. Next, we derived the upper limits on the rate density of the GWHEN sources as a function of the total bolometric energy emitted in HEN ($E_\nu$), separately for each merger type.

In Table \ref{tab:my_label}, we provide the 90\% upper limits on the rate density of GWHEN events we obtained as a function of $E_\nu$, assuming isotropic emission. They are the first limits put on the population of merging compact binaries. We also show the latest estimates of the rate densities of each merger type after O3~\citep{PhysRevX.13.011048}. A GWHEN limit below these rate densities, for example for $E_\nu\gtrsim10^{52}$--$10^{54}$~erg, implies that only a portion of that merger type can be emitting HEN at $E_\nu$, assuming all the emissions have the same $E_\nu$. If beamed emission are assumed instead of isotropic emission, obtained upper limits become higher and do not constrain the presence of such emissions from the population. Energies around as $10^{52}$--$10^{54}$~erg are too large for the expected total energies emitted in HEN for such mergers. Moreover, these emissions are expected to be beamed. Hence, the first limits on the rate density of joint GW and HEN emission from the population of merging compact binaries are not too useful except for constraining very optimistic isotropic emission models. The improvements on these limits can come with increasing detector sensitivities and with extended datasets.
\begin{table}
    \centering
    \begin{tabular}{|c|c|c|c|c|}
    \hline
         & $E_\nu=10^{50}$~erg & $E_\nu=10^{52}$~erg  & $E_\nu=10^{54}$~erg & GW event rate density\\
         \hline
        BNS & 50000 & 890 & 348& 10--1700 \\
        \hline
        NSBH & 20000 & 290 & 63 & 7.8--140\\
        \hline
        BBH & 19000 & 220 & 14 & 17.9--44\\
        \hline
    \end{tabular}
    \caption{90\% upper limits on the GWHEN event rate densities assuming isotropic emissions for different emitted total bolometric HEN energies. For comparison, the latest 90\% credible intervals on the GW rate densities~\citep{PhysRevX.13.011048} are given in the last column. Units are in Gpc$^{-3}$yr$^{-1}$.}
    \label{tab:my_label}
\end{table}

\bibliographystyle{ICRC}
\bibliography{references}

\clearpage

\section*{Full Author List: IceCube Collaboration}

\scriptsize
\noindent
R. Abbasi$^{16}$,
M. Ackermann$^{63}$,
J. Adams$^{17}$,
S. K. Agarwalla$^{39,\: {\rm a}}$,
J. A. Aguilar$^{10}$,
M. Ahlers$^{21}$,
J.M. Alameddine$^{22}$,
S. Ali$^{35}$,
N. M. Amin$^{43}$,
K. Andeen$^{41}$
% ...
\\
\\
% ...
$^{10}$ Universit{\'e} Libre de Bruxelles, Science Faculty CP230, B-1050 Brussels, Belgium \\
% ...
$^{16}$ Department of Physics, Loyola University Chicago, Chicago, IL 60660, USA \\
$^{17}$ Dept. of Physics and Astronomy, University of Canterbury, Private Bag 4800, Christchurch, New Zealand \\
% ...
$^{21}$ Niels Bohr Institute, University of Copenhagen, DK-2100 Copenhagen, Denmark \\
$^{22}$ Dept. of Physics, TU Dortmund University, D-44221 Dortmund, Germany \\
% ...
$^{35}$ Dept. of Physics and Astronomy, University of Kansas, Lawrence, KS 66045, USA \\
% ...
$^{39}$ Dept. of Physics and Wisconsin IceCube Particle Astrophysics Center, University of Wisconsin{\textemdash}Madison, Madison, WI 53706, USA \\
% ...
$^{41}$ Department of Physics, Marquette University, Milwaukee, WI 53201, USA \\
% ...
$^{43}$ Bartol Research Institute and Dept. of Physics and Astronomy, University of Delaware, Newark, DE 19716, USA \\
% ...
$^{63}$ Deutsches Elektronen-Synchrotron DESY, Platanenallee 6, D-15738 Zeuthen, Germany \\
$^{\rm a}$ also at Institute of Physics, Sachivalaya Marg, Sainik School Post, Bhubaneswar 751005, India \\
% ...

\subsection*{Acknowledgments}

\noindent
The authors gratefully acknowledge the support from the following agencies and institutions:
USA {\textendash} U.S. National Science Foundation-Office of Polar Programs,
U.S. National Science Foundation-Physics Division,
U.S. National Science Foundation-EPSCoR,
U.S. National Science Foundation-Office of Advanced Cyberinfrastructure,
Wisconsin Alumni Research Foundation,
Center for High Throughput Computing (CHTC) at the University of Wisconsin{\textendash}Madison,
Open Science Grid (OSG),
Partnership to Advance Throughput Computing (PATh),
Advanced Cyberinfrastructure Coordination Ecosystem: Services {\&} Support (ACCESS),
Frontera and Ranch computing project at the Texas Advanced Computing Center,
U.S. Department of Energy-National Energy Research Scientific Computing Center,
Particle astrophysics research computing center at the University of Maryland,
Institute for Cyber-Enabled Research at Michigan State University,
Astroparticle physics computational facility at Marquette University,
NVIDIA Corporation,
and Google Cloud Platform;
Belgium {\textendash} Funds for Scientific Research (FRS-FNRS and FWO),
FWO Odysseus and Big Science programmes,
and Belgian Federal Science Policy Office (Belspo);
Germany {\textendash} Bundesministerium f{\"u}r Forschung, Technologie und Raumfahrt (BMFTR),
Deutsche Forschungsgemeinschaft (DFG),
Helmholtz Alliance for Astroparticle Physics (HAP),
Initiative and Networking Fund of the Helmholtz Association,
Deutsches Elektronen Synchrotron (DESY),
and High Performance Computing cluster of the RWTH Aachen;
Sweden {\textendash} Swedish Research Council,
Swedish Polar Research Secretariat,
Swedish National Infrastructure for Computing (SNIC),
and Knut and Alice Wallenberg Foundation;
European Union {\textendash} EGI Advanced Computing for research;
Australia {\textendash} Australian Research Council;
Canada {\textendash} Natural Sciences and Engineering Research Council of Canada,
Calcul Qu{\'e}bec, Compute Ontario, Canada Foundation for Innovation, WestGrid, and Digital Research Alliance of Canada;
Denmark {\textendash} Villum Fonden, Carlsberg Foundation, and European Commission;
New Zealand {\textendash} Marsden Fund;
Japan {\textendash} Japan Society for Promotion of Science (JSPS)
and Institute for Global Prominent Research (IGPR) of Chiba University;
Korea {\textendash} National Research Foundation of Korea (NRF);
Switzerland {\textendash} Swiss National Science Foundation (SNSF).

This material is based upon work supported by NSF’s LIGO Laboratory, which is a major facility fully funded
by the National Science Foundation. The authors also gratefully acknowledge the support of the Science
and Technology Facilities Council (STFC) of the United Kingdom, the Max-Planck-Society (MPS), and the
State of Niedersachsen/Germany for support of the construction of Advanced LIGO and construction and
operation of the GEO 600 detector. Additional support for Advanced LIGO was provided by the Australian
Research Council. The authors gratefully acknowledge the Italian Istituto Nazionale di Fisica Nucleare
(INFN), the French Centre National de la Recherche Scientifique (CNRS) and the Netherlands Organization for Scientific Research (NWO) for the construction and operation of the Virgo detector and the creation
and support of the EGO consortium. The authors also gratefully acknowledge research support from these
agencies as well as by the Council of Scientific and Industrial Research of India, the Department of Science
and Technology, India, the Science \& Engineering Research Board (SERB), India, the Ministry of Human
Resource Development, India, the Spanish Agencia Estatal de Investigacion (AEI), the Spanish Ministerio ´
de Ciencia, Innovacion y Universidades, the European Union NextGenerationEU/PRTR (PRTR-C17.I1), the ´
ICSC - CentroNazionale di Ricerca in High Performance Computing, Big Data and Quantum Computing,
funded by the European Union NextGenerationEU, the Comunitat Autonoma de les Illes Balears through `
the Conselleria d’Educacio i Universitats, the Conselleria d’Innovaci ´ o, Universitats, Ci ´ encia i Societat Dig- `
ital de la Generalitat Valenciana and the CERCA Programme Generalitat de Catalunya, Spain, the Polish
National Agency for Academic Exchange, the National Science Centre of Poland and the European Union
- European Regional Development Fund; the Foundation for Polish Science (FNP), the Polish Ministry of
Science and Higher Education, the Swiss National Science Foundation (SNSF), the Russian Science Foundation, the European Commission, the European Social Funds (ESF), the European Regional Development
Funds (ERDF), the Royal Society, the Scottish Funding Council, the Scottish Universities Physics Alliance,
the Hungarian Scientific Research Fund (OTKA), the French Lyon Institute of Origins (LIO), the Belgian
Fonds de la Recherche Scientifique (FRS-FNRS), Actions de Recherche Concertees (ARC) and Fonds ´
Wetenschappelijk Onderzoek - Vlaanderen (FWO), Belgium, the Paris ˆIle-de-France Region, the National
Research, Development and Innovation Office of Hungary (NKFIH), the National Research Foundation
of Korea, the Natural Sciences and Engineering Research Council of Canada (NSERC), the Canadian
Foundation for Innovation (CFI), the Brazilian Ministry of Science, Technology, and Innovations, the International Center for Theoretical Physics South American Institute for Fundamental Research (ICTP-SAIFR),
the Research Grants Council of Hong Kong, the National Natural Science Foundation of China (NSFC), the
Israel Science Foundation (ISF), the US-Israel Binational Science Fund (BSF), the Leverhulme Trust, the
Research Corporation, the National Science and Technology Council (NSTC), Taiwan, the United States
Department of Energy, and the Kavli Foundation. The authors gratefully acknowledge the support of the
NSF, STFC, INFN and CNRS for provision of computational resources.
This work was supported by MEXT, the JSPS Leading-edge Research Infrastructure Program, JSPS Grantin-Aid for Specially Promoted Research 26000005, JSPS Grant-in-Aid for Scientific Research on Innovative
Areas 2905: JP17H06358, JP17H06361 and JP17H06364, JSPS Core-to-Core Program A. Advanced Research Networks, JSPS Grants-in-Aid for Scientific Research (S) 17H06133 and 20H05639, JSPS Grant-inAid for Transformative Research Areas (A) 20A203: JP20H05854, the joint research program of the Institute
for Cosmic Ray Research, the University of Tokyo, the National Research Foundation (NRF), the Computing
Infrastructure Project of Global Science experimental Data hub Center (GSDC) at KISTI, the Korea Astronomy and Space Science Institute (KASI), the Ministry of Science and ICT (MSIT) in Korea, Academia Sinica
(AS), the AS Grid Center (ASGC) and the National Science and Technology Council (NSTC) in Taiwan under grants including the Science Vanguard Research Program, the Advanced Technology Center (ATC) of
NAOJ, and the Mechanical Engineering Center of KEK.

\end{document}